\documentclass[12pt,preprint]{aastex}

\usepackage{graphicx}

\shorttitle{GRTidalDisruption}
\shortauthors{Dai et al.}

\begin{document}

\title{The Impact of Bound Stellar Orbits and General Relativity \\
on the Temporal Behavior of Tidal Disruption Flares}

\author{Lixin Dai}
\affil{Departamento de Astronomia, Universidad de Chile, Santiago, Chile;  \\ Department of Astronomy, Yale University, New Haven, CT 06511, USA\\ janedai@das.uchile.cl}

\author{ Andres Escala}
\affil{Departamento de Astronomia, Universidad de Chile, Santiago, Chile}

\author{ Paolo Coppi}
\affil{Department of Astronomy, Yale University, New Haven, CT 06511}

\begin{abstract}
We have carried out general relativistic particle simulations of stars tidally disrupted by massive black holes. When a star is disrupted in a bound orbit with moderate eccentricity instead of a parabolic orbit, the temporal behavior of the resulting stellar debris changes qualitatively. The debris is initially all bound, returning to pericenter in a short time $\sim$ the original stellar orbital timescale. The resulting fallback rate can thus be much higher than the Eddington rate. Furthermore if the star is disrupted close to the hole, in a regime where general relativity is important, the stellar and debris orbits display general relativistic precession. Apsidal precession can make the debris stream cross itself after several orbits, likely leading to fast debris energy dissipation. If the star is disrupted in an inclined orbit around a spinning hole, nodal precession reduces the probability of self-intersection, and circularization may take many dynamical timescales, delaying the onset of flare activity. An examination of the particle dynamics suggests that quasi-periodic flares with short durations, produced when the center of the tidal stream passes pericenter, may occur in the early-time light curve. The late-time light curve may still show power-law behavior which is generic to disk accretion processes. The detection triggers for future surveys should be extended to capture such ``non-standard" short-term flaring activity before the  event enters the asymptotic decay phase, as this activity is likely to be more sensitive to physical parameters such as the black hole spin.
\end{abstract}


\section{Introduction}
Tidal disruption (TD) of stars by massive black holes (MBHs) has been theoretically predicted since the 1970s \citep[e.g.,][]{Hills75, Lacy82, Carter83}. The detection of tidal disruption events (TDEs) can be used to probe dormant MBHs in distant galaxies. The white dwarf TD, if ever detected, would be evidence for the existence of intermediate-mass black holes (IMBHs). Today $\sim 20$ TD candidates \citep[and the references therein]{Komossa99a, Komossa99b, Gezari09, Komossa12} have been observed, including two possible relativistic TD flares \citep[e.g.,][]{Bloom11, Cenko12} with X-ray quasi-periodic oscillations reported for the \objectname{Sw 1644+57} event \citep{Reis12}. Upcoming time-domain surveys, e.g., by the Large Synoptic Survey Telescope (LSST), will increase the TDE sample size by a factor of 100 \citep{Strubbe09}. It is thus important to study the temporal behavior of TD flares to detect TDEs more effectively and probe MBH space-time and accretion physics.

\citet{Rees88} has presented a canonical picture of TDEs which is currently used to identify TDE candidates. Here a solar-type star is scattered from faraway into the loss cone of an MBH, putting the star into a parabolic orbit. If the orbital pericenter distance $R_p$ is smaller than the tidal radius $R_T$, the distance where the tidal force from the MBH equals the stellar self-gravity, the star is torn apart by tidal forces in a single flyby. The tidal radius is:
\begin{equation}
R_T = \eta \times R_\star \ (M/m_\star)^{1/3},
\end{equation}
where $M$ is the MBH mass, $m_\star$ and $R_\star$ are the stellar mass and radius, and $\eta$ is a constant of order unity determined by the stellar model (hereafter we take $\eta = 1$). After disruption, stellar debris initially follows ballistic orbits determined by the distribution of debris binding energy at the time of disruption. Half of the debris escapes in hyperbolic orbits, and the other half remains bound to the MBH. If the stellar mass per binding energy, $dm / dE$, is constant, the rate of debris mass orbiting back to pericenter (the so-called ``fallback'' rate) goes as a power law with time $\sim t^{-5/3}$ \citep{Phinney89, Evans89, Lodato09}. The hydrodynamical interaction of the debris such as tidal stream collisions, tidal compression, and shock formation should lead to the circularization of debris orbits and the eventual formation of an accretion disk \citep{Kochanek94, Guillochon13}. As the typical fallback timescale (years) is long compared to the viscous timescale of a standard accretion disk, the accretion rate likely follows the fallback rate, at least at late times \citep{Rees88, Ulmer99}. The result is a luminous electromagnetic flare, likely peaked in the UV/X-rays, which also decays with $t^{-5/3}$ \citep{Strubbe09, Ayal00}.

As recognized in \citet{Rees88}, this TD picture glosses over many important details such as the possibility of deep encounters (i.e., the penetration parameter $\beta = R_T / R_p > 1$), variations in the structure of the star, and the change of debris binding energy by tidal compression. More realistic treatment of these details can lead to different observational predictions  \citep[e.g.,][]{Rosswog09, Guillochon09, Guillochon12, Stone13}. Among the possible complications, we focus on two of them---the parabolic nature of the stellar orbits and the effects of general relativity (GR) on the debris orbits.

1. The star is conventionally assumed to approach the MBH in a parabolic orbit (the orbital eccentricity $e=1$), since it loses energy and angular momentum slowly through many weak scatterings faraway from the hole till it just becomes marginally bound. However, there are several mechanisms that can produce eccentric TDEs with $e<1$, as summarized in \citet{Hayasaki13}. For example, one strong encounter with a massive perturber such as a giant molecular cloud (GMC) or an IMBH could remove enough angular momentum to put the star in a bound orbit (e.g., see the review by \citet{Alexander12}). The overall TDE rate can be boosted by as much as three orders of magnitude when an IMBH is present \citep{Chen09, Chen11}. GMCs can have complicated interactions with MBHs and their stellar cusps, leading to the production of stellar disks near MBHs \citep[e.g.,][]{Yusef12}. It is not clear yet what the actual TD event rate is for $e<1$. However, the fact that we see many young, bound stars near the Galactic center (e.g., the ``S-stars''  found by \citealt{Ghez05} and \citealt{Gillessen09}) suggests that there are ways to put stars in orbits quite close to MBHs, lessening the requirement that TDEs are likely only to happen for near-parabolic trajectories. In this regard, we know that even an $e\sim 0.99$ orbit will lead to fallback rate and tidal stream orbits that are qualitatively different from the $e = 1$ case, as we show in this Letter.

2. Even for the parabolic case, the star can be disrupted close to the MBH, at a few gravitational radii $R_g = G M / c^2$ (where $G$ is the gravitational constant, and $c$ is the speed of light). If the dimensionless TD radius
\begin{equation}
\tilde{R}_T = R_T / R_g \propto M^{-2/3}{\bar{\rho}_\star}^{-1/3},
\end{equation}
where $\bar{\rho}_\star$ is the mean stellar density, approaches order unity, GR effects become significant. The parameter space for relativistic TDEs is important: a white dwarf can only be disrupted by IMBHs within $\sim 10 R_g$; the TD by higher mass black holes with masses of $\sim10^7 -10^8 M_\odot$ and even rapidly spinning ones of $10^9 M_\odot$ \citep{Kesden12a}, if happens, must be relativistic, unless the star is a giant with an extended envelope.

If a star is disrupted in a bound orbit at $\tilde{R}_T < 10$, the behavior of the debris orbits, the debris circularization mechanism, and the accretion and radiation physics could all be different from the standard scenario. \citet{Rasio05}, \citet{Kesden12a, Kesden12b}, and \citet{Haas12} calculated some features of stars disrupted in GR parabolic orbits. \citet{Hayasaki13} carried out hydrodynamical simulations of a star disrupted in a bound orbit, using a pseudo-Newtonian potential which is a good approximation to the Schwarzschild potential for $e \sim 1$. In this Letter, we study the disruption of a star in a bound orbit by a supermassive black hole (SMBH), and extend the previous work by using a suite of particle simulations in the full Kerr metric. Our simulation does not include hydrodynamical effects, but can explore more parameter space due to the fast computation time. In Section 2, we introduce our code and the setup of the problem. In Section 3, we examine the importance of bound stellar orbits and GR during the initial phase of the tidal debris orbits. In the last section, we summarize the results and speculate on the observational consequences.

\section{Methodology}

For all the results in this Letter, we simulated the debris orbits of a lower main-sequence star with mass $\sim 0.3 M_\odot$ when disrupted by a $10^7 M_\odot$ MBH. Using Equation (2), $\tilde{R}_T \sim 7.1 R_g$, which is in the strong relativity regime. We sampled the star with $10^6$ identical mass particles, drawn from a density distribution calculated from the Lane-Emden equation with a polytropic index of $\gamma = 5/3$.

We assume that all the star particles move as a single unit with the center of the mass before disruption. There is no closed eccentric orbit in GR, but the stellar orbit can be approximated by an ellipse with apsidal and nodal precession. The eccentricity can be defined by $e = (R_a-R_p)/(R_a+R_p)$, where $R_a$ is the apocenter distance. After disruption, each debris particle travels on its own geodesic calculated using the position and momentum of the particle at disruption. The effects of tidal compression before disruption and self-gravity are ignored in these simulations. However, as $\beta = 1$ in all our simulations, tidal compression is not strong near pericenter \citep{Carter83}. The fifth-order Runge--Kutta algorithm is used to integrate debris orbits under the Kerr metric.  

\section{Results}

\subsection{Eccentric Orbit: Small Spread in Fallback Time}

\begin{figure}
\centering
 \includegraphics[width=6in]{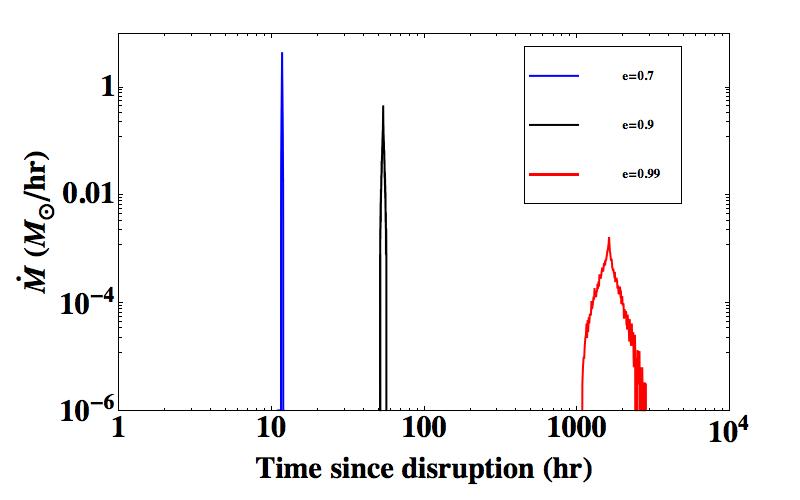}
\caption{The fallback rate as a function of time after disruption for a $\sim 0.3 M_\odot$ lower main-sequence star disrupted by a $10^7 M_\odot$ Schwarzschild MBH. The star is put in bound orbits with the same $R_p \sim 7.1 R_g$ but different eccentricities $e$. The spread of the fallback time and peak fallback rate are very sensitive to $e$. }
\label{Eccentricities}
\end{figure}

If a star is disrupted in a bound orbit, the center of the star has a negative binding energy. Therefore, more than half of the debris remains bound and eventually accretes onto the MBH. For small enough eccentricity all debris can be bound \citep{Hayasaki13}, and quickly returns to pericenter with very small spread of fallback time. Fig. \ref{Eccentricities} shows the fallback rate as a function of eccentricity. The spread of fallback time decreases from $\sim 1$ yr ($e=1$) to weeks ($e=0.99$), days ($e=0.9$), or even hours or minutes ($e=0.7$), leading to a very high mass fallback rate which does not decay as $t^{-5/3}$.

How the fallback rate translates to the accretion rate or flare luminosity is still unclear. Recently, \citet{Guillochon13} has carried out hydrodynamical simulations of the evolution of the tidal debris from a star disrupted in a parabolic orbit by a SMBH, and found that dissipation through shocks likely cannot circularize tidal debris fast enough to produce the observed flare luminosities. In relativistic eccentric TDEs, debris particles tend to travel together and do not go out to large distances, suggesting that hydrodynamical dissipation based on shocks at the ``nozzle'', shearing, and compression may be even weaker. However, one important effect speculated upon by \citet{Guillochon13}  --- self-intersection of the tidal stream due to apsidal GR precession --- may even work better. We will investigate this next. 
 
\subsection{Schwarzschild MBH: Self-intersection of the Tidal Stream}

\begin{figure}
\centering
 \includegraphics[width=6in]{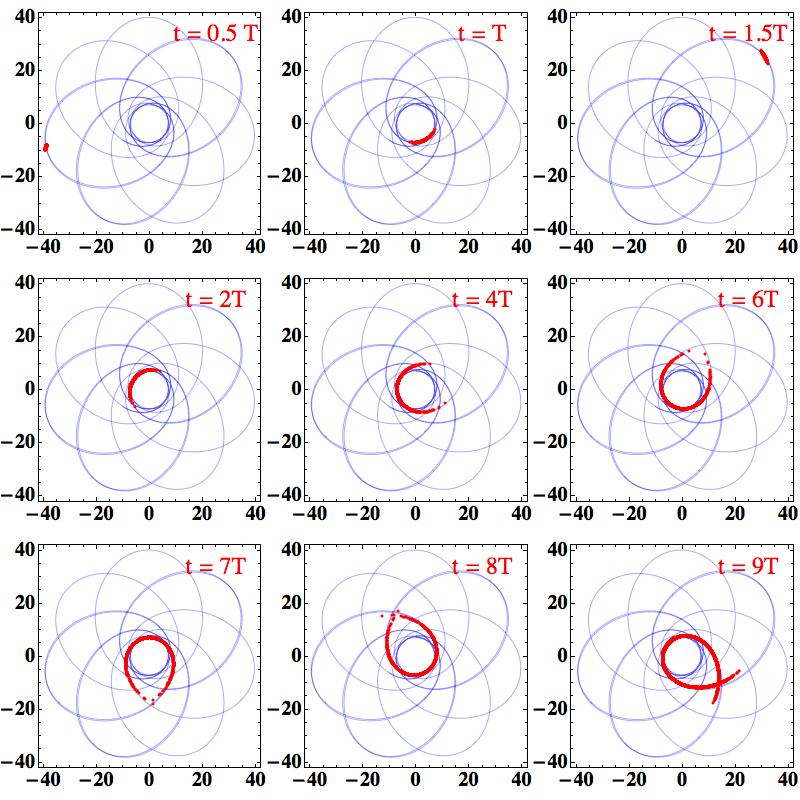}
\caption{Snapshots of the tidal stream of a $\sim 0.3 M_\odot$ lower main-sequence star disrupted by a Schwarzschild $10^7 M_\odot$ black hole. The geodesic of the center of the star has $e=0.7$, $R_p \sim 7 R_g$, and $R_a \sim 40 R_g$, shown by the blue curve. The red points show the locations of debris particles at different epochs. $T$ is the orbital time of the original stellar orbit. The star is disrupted at $R_p$, so at $t=T/2$ the debris goes to apocenter for the first time and at $t=T$ returns to pericenter for the first time. The length of the tidal stream grows with time. Around the ninth orbit, the stream starts to cross itself.}
\label{SelfCrossing}
\end{figure}

Several orbits after disruption, the tidal stream is tangentially elongated and almost resides on the geodesic of the original center of star (Fig. \ref{SelfCrossing}). In the Schwarzschild metric, the geodesic is subject to apsidal precession. By the time the center of the mass returns to pericenter, the ``head'' of the stream may have precessed enough so it can cross the``tail'' of the stream. We show an example in Fig. 3 where the stream intersects with itself in the ninth orbit when $e=0.7$. If the same star is disrupted in an orbit with $e = 0.9$, self-intersection can even happen in the first orbit due to the larger orbital period spread of the debris particles. Therefore, if we increase the eccentricity, the possibility of self-intersection increases.

As we can see in Fig.  \ref{SelfCrossing}, when self-intersection happens, the collision angle can be large. Therefore, strong energy dissipation due to heating and shocks will possibly circularize debris in several orbits. Indeed, the hydrodynamical simulation of \citet{Hayasaki13} shows that the stream trajectory has no significant deviation from geodesics until self-intersection happens. This suggests that self-intersection is the main mechanism for debris circularization in eccentric TDEs. If the circularization does happen in several orbits, the viscous timescale for a standard accretion disk is much longer than circularization timescale. This means that the debris will accumulate and probably form a thick disk, with an accretion rate not following the fallback rate.

\subsection{Kerr MBH: Delay of the Tidal Stream Self-intersection}
\begin{figure}
\centering
 \includegraphics[width=6in]{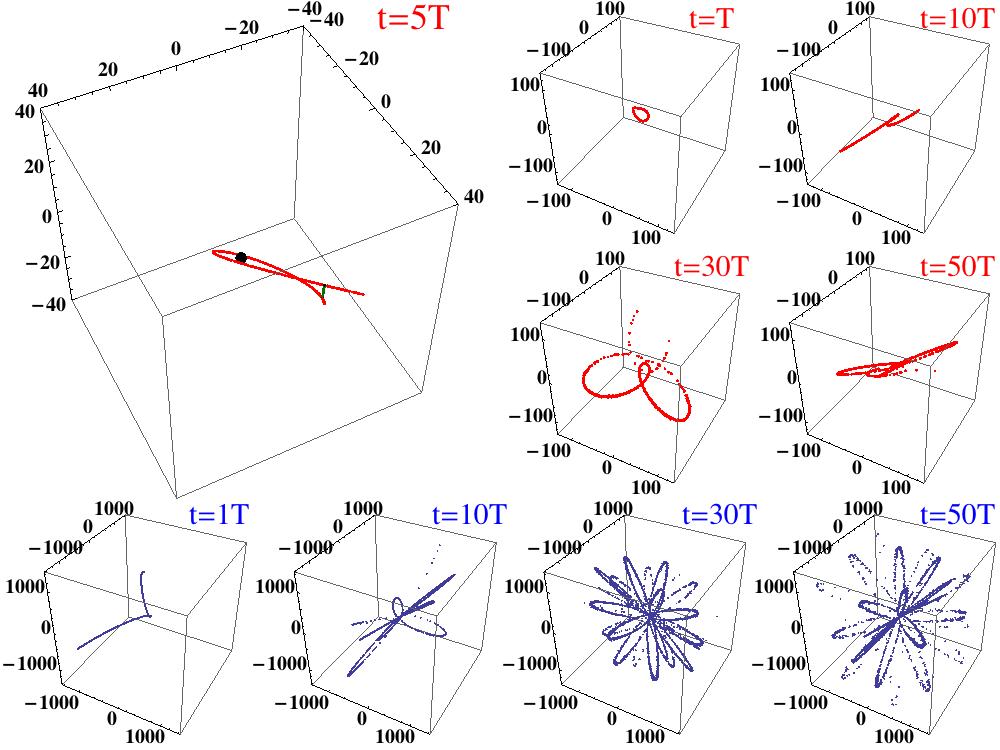}
\caption{Snapshots of the tidal stream of a $\sim 0.3 M_\odot$ lower main-sequence star disrupted in an eccentric orbit with $e = 0.9$, $i = 45^\circ$ (red dots) or $e = 0.99, i = 45^\circ$ (blue dots) around a Kerr black hole of $M=10^7 M_\odot$ and $\tilde{S} = 0.5$. The zoomed-in snapshot at $t=5T$ shows how the tidal stream misses itself due to orbital plane precession. The black dot there represents the MBH, and the green line shows the distance of closest approach $d$. }
\label{Kerr}
\end{figure}

MBHs may well be spinning, and there is no reason to assume that the stellar orbital plane is perpendicular to the MBH spin axis (e.g., the ``S-stars'' in the Galactic center). If the two are not perpendicular, then nodal precession dominated by the Lense--Thirring effect will shift the orbital plane and delay tidal stream self-intersection. We show in Fig. \ref{Kerr} the snapshots of the tidal stream at different epochs, after the star is disrupted in an inclined orbit by a Kerr hole. The initial phase of the tidal stream looks similar to the Schwarzschild one, but as can be seen in the zoomed snapshot that the stream misses itself by a small distance $d$ due to nodal precession. 

\begin{figure}
\centering
 \includegraphics[width=3.5in]{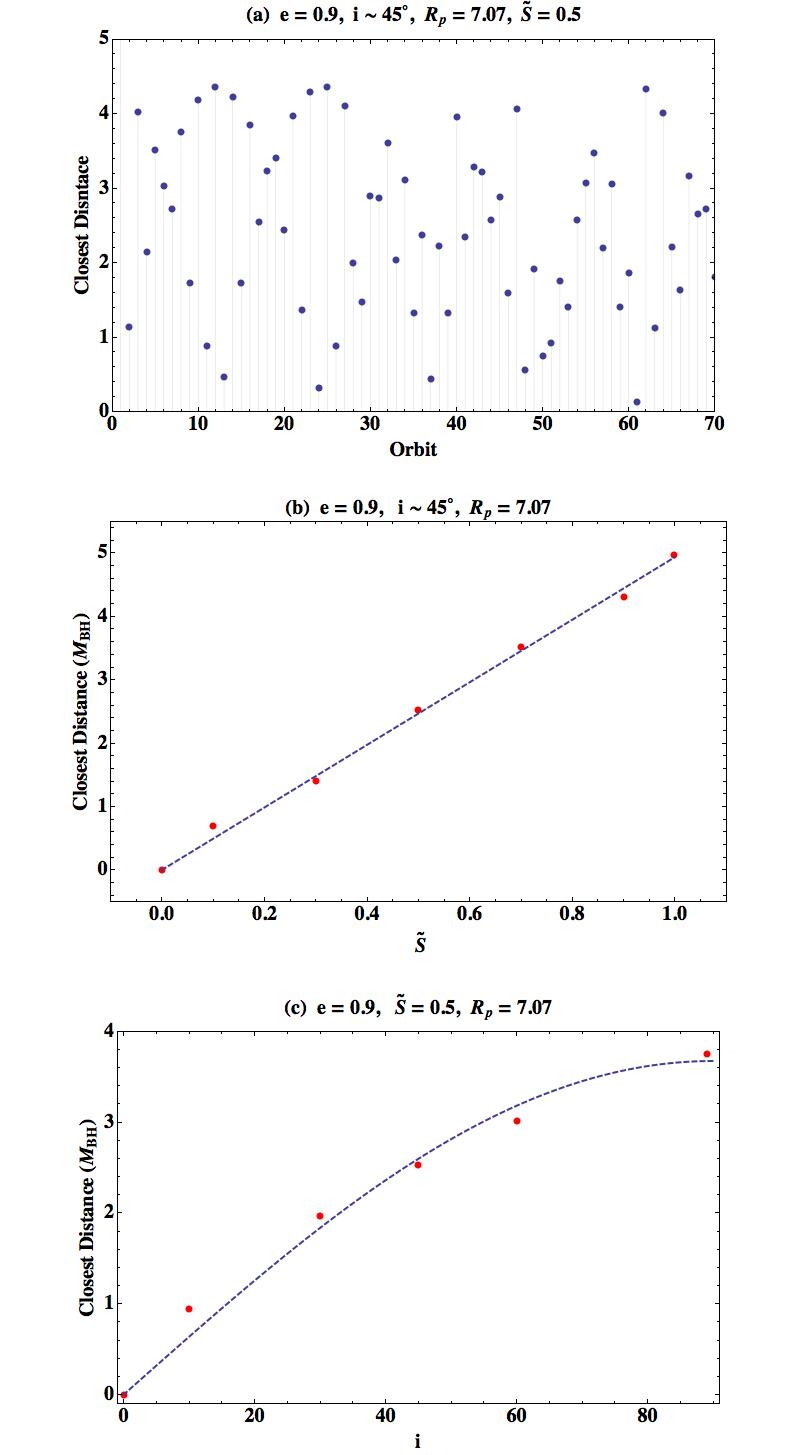}
\caption{Panel(a) shows the value of $d$ when the stream center returns to pericenter for the first 70 times. The initial stellar orbit has $e = 0.9, i  = 45^\circ$, and the MBH has $\tilde{S} = 0.5$. This distance varies but is at least one order above the tidal stream thickness $\sim 0.04 R_g$. Panel(b) shows how the averaged $d$ over the first 70 orbits  changes with $\tilde{S}$ (ranging from 0 to 0.998) for a fixed stellar orbit with $e=0.9, i=45^\circ$. The blue dashed line is a linear fit to the simulation results (red points). Panel(c) shows how the orbit-averaged $d$ changes when we fix $\tilde{S} = 0.5$ but change $i$ instead. The blue dashed line is a sine fit to the simulation results (red points). }
 \label{Distance}
\end{figure}

This distance of closest approach $d$ between the stream is plotted in Fig. \ref{Distance}(a) for the first 70 times that the center of mass returns to pericenter. Due to the advance of pericenter on the orbital plane, $d$ is different each time. In order to see if self-intersection will happen, we need to compare $d$ to the thickness $h$ of the tidal stream, which is not obtainable from our simulation since it does not include hydrodynamical effects. An order-of-magnitude calculation by \citet{Kochanek94} has $h\propto r^{1/4}$, where $r$ is the distance from the hole. This result is confirmed by \citet{Guillochon13} for the first orbit after the disruption of a star in a parabolic trajectory. The simulation of an $e = 0.9, i=45^{\circ}$ ($i$ is the orbital inclination angle) stellar orbit shows that the closest approach between the tidal stream happens at $r_c \sim 29.4 R_g$,  so given the original stellar size $R_\star \sim 0.3 R_\odot$ at $R_p \sim 7.1R_g$, we get $h(r_c) \sim 0.04 R_g$. Therefore, only if $ d \lesssim 0.04 R_g$, there is strong self-intersection of the tidal stream. This does not happen in these 70 orbits.

The value of $d$ depends on several parameters including the orbital size, eccentricity $e$, inclination angle $i$, and the MBH spin $\tilde{S}$, since the Lense-Thirring precession rate depends on these parameters. We show that the orbit-averaged $d$ changes linearly with $\tilde{S}$ and sinusoidally with $i$ in Fig. \ref{Distance}(b) and (c). This dependence agrees with the Lense--Thirring precession formula. Thus, if the estimation of the debris thickness is correct, self-intersection of the tidal stream due to GR precession can only happen quickly if $\tilde{S} \sim 0$ or $i\sim 0$.

\begin{figure}
\centering
 \includegraphics[width=6in]{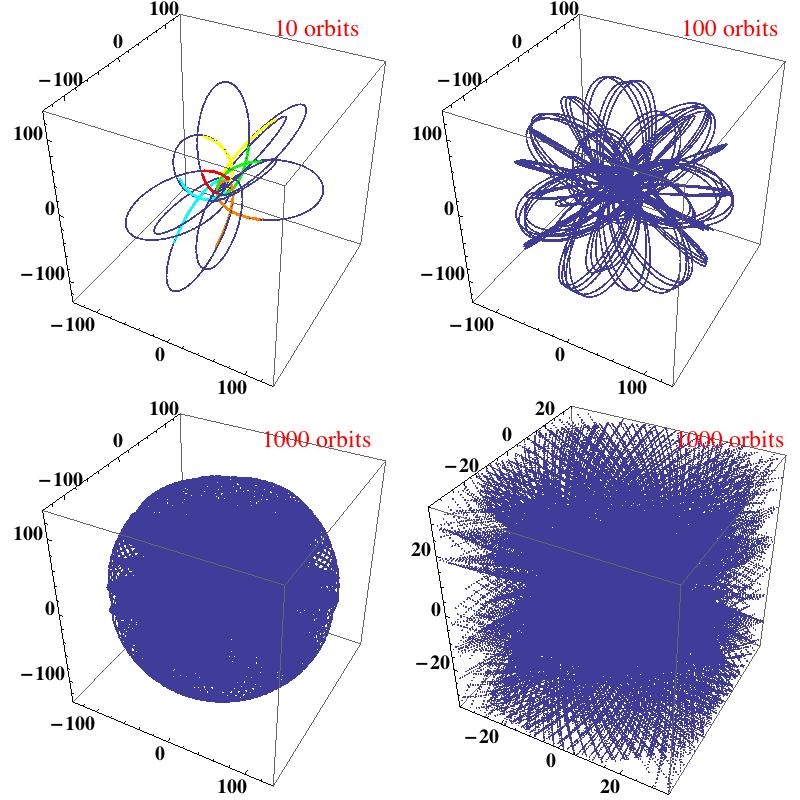}
 \caption{The geodesic for the center of a star in an eccentric orbit with $e = 0.9, i = 45^\circ$ around a Kerr black hole with $M=10^7 M_\odot$ and $\tilde{S} = 0.5$. $R_p \sim 7.1 R_g$, and $R_a \sim 134.3 R_g$. We show the first 10, 100, 1000 orbits and a zoomed-in of the 1000 orbits case. For the panel showing 10 orbits, we also overplot the locations of the tidal streams at $t = T$ (red), $t = 2T$ (green), $t = 3T$ (orange), $t = 4T$ (cyan), and $t = 5T$ (yellow). All tidal streams tend to align with the geodesic curve.}
 \label{KerrGeodesic}
\end{figure}

Following the tidal stream evolution for $>100$ orbits is difficult, because the stream becomes very long and has a complicated shape thus requiring too many particles to sample it well. 
However, the tidal stream tends to lie on the geodesic of the center of the star (see Fig. \ref{KerrGeodesic}). In order to get a sense of how the tidal stream fills the space around the hole and thus how long it takes for the self-intersection to happen, in Fig. \ref{KerrGeodesic} we plot the first 10, 100, and 1000 orbits of the center of the star. If the tidal stream maintains ballistic behavior after 1000 orbits, it forms a gas ``ball'' around the hole, reminiscent of the scenarios in \citet{Loeb97} and \citet{Ulmer98}, although this sphere of gas is not supported by pressure.

\section{Summary and Discussion}

In conclusion, if a star is tidally disrupted in a bound orbit with moderate eccentricity ($e\lesssim0.99$)  all the debris remains bound, and returns to pericenter in a spread of time much shorter (up to a factor of 10000) than the typical fallback time of a star disrupted in a parabolic orbit. The fallback rate is very large and can even be super-Eddington, and does not decay with $t^{-5/3}$. It may be hard for the debris energy to dissipate unless the star is disrupted close to the MBH and apsidal precession can make tidal stream intersect with itself. However, if the MBH is spinning and if its spin axis is not perpendicular to the stellar orbital plane, nodal precession can greatly delay the stream self-intersection if the stream is as thin as seen in recent numerical simulations. This would delay the energy dissipation and the appearance of the flare. If no hydrodynamical interactions destroy the stream, a thick ``ball'' of precession gas  is likely to form around the MBH. Note that as a function of distance away from the MBH nodal precession rate drops faster than apsidal precession rate, so there may be a regime where nodal precession does not delay the stream self-intersection.

Hydrodynamical simulations need to be conducted to accurately calculate the debris accretion and the light curve for eccentric TDEs. However, the dynamics of the debris orbits seen here suggests that eccentric relativistic TDEs can look quite different from standard ones. The total duration of the flare could be much shorter. In the early phase of the light curve, short bursts with durations ($\sim 1$--$10$ hr) imprinted by the spread of debris fallback time may be observed quasi-periodically with period $P$ ($\sim 1$--$10$ days) which is roughly the stellar orbital period, especially if stream self-intersection is important. Nodal precession will change the light curve, likely lowering the magnitude of these short bursts, but also providing us with the probe of the black hole spacetime. We know that \citet{Levan11} found fast large-amplitude variability in the initial light curve of \objectname{Sw 1644+57}, and four short flares were observed with a quasi-period of $\sim$10 hr. The very late time light curve might still exhibit power-law behavior as a result of the disk accretion processes. Therefore, the early-time TD light curve contains more information on the MBH and stellar orbital parameters, and thus should be monitored more closely. 

The rate of eccentric relativistic TDEs may be low, but the observation of even just one or two of them can greatly improve our understanding on the dynamics and accretion around MBHs. In the upcoming decade, the launch of LSST and other telescopes will greatly enlarge the TDE sample size, and we will likely observe such non-standard TDEs if the appropriate trigger is implemented to catch them.

\acknowledgments
We are grateful to Roger Blandford and Priya Natarajan for discussions and encouragement. The stellar data is kindly provided by Peter Eggleton. We thank the anonymous referee for very useful suggestions and comments. We would also like to thank Pau Amaro-Seoane, Xian Chen, James Guillochon, Michael Kesden, Jonathan McKinney, David Merritt, Cole Miller, Gabe Perez-Giz, Enrico Ramirez-Ruiz, Nick Stone, and Alexander Tchekhovskoy for useful conversations. L.D. acknowledges the support from the Financiamiento Basal Grant PFB 06 and the Yale-Universidad de Chile Joint Program. A.E. acknowledges partial support from the Financiamiento Basal Grant PFB 06, FONDECYT Grant 1130458 and Anillo de Ciencia y Tecnologia Grant ACT1101.

\clearpage

\end{document}